\documentclass[english,apl, twocolumn, superscriptaddress]{revtex4}
\usepackage[T1]{fontenc}
\usepackage[latin9]{inputenc}
\usepackage{graphicx}

\makeatletter
\@ifundefined{textcolor}{}
{%
 \definecolor{BLACK}{gray}{0}
 \definecolor{WHITE}{gray}{1}
 \definecolor{RED}{rgb}{1,0,0}
 \definecolor{GREEN}{rgb}{0,1,0}
 \definecolor{BLUE}{rgb}{0,0,1}
 \definecolor{CYAN}{cmyk}{1,0,0,0}
 \definecolor{MAGENTA}{cmyk}{0,1,0,0}
 \definecolor{YELLOW}{cmyk}{0,0,1,0}
}

\makeatother

\usepackage{babel}
\begin{document}

\title{All-Optical Depletion of Dark Excitons from a Semiconductor Quantum Dot}

\author{E. R. Schmidgall}

\affiliation{The Physics Department and the Solid State Institute, Technion-Israel
Institute of Technology, Haifa 32000, Israel}

\author{I. Schwartz}

\affiliation{The Physics Department and the Solid State Institute, Technion-Israel
Institute of Technology, Haifa 32000, Israel}

\author{D. Cogan}

\affiliation{The Physics Department and the Solid State Institute, Technion-Israel
Institute of Technology, Haifa 32000, Israel}

\author{L. Gantz}

\affiliation{The Physics Department and the Solid State Institute, Technion-Israel
Institute of Technology, Haifa 32000, Israel}

\affiliation{Department of Electrical Engineering, Technion-Israel Institute of
Technology, Hafia 32000, Israel}

\author{T. Heindel}

\affiliation{Institute of Solid State Physics, Technische Universit{\"a}t Berlin, 10623 Berlin, Germany}

\author{S. Reitzenstein}

\affiliation{Institute of Solid State Physics, Technische Universit{\"a}t Berlin, 10623 Berlin, Germany}

\author{D. Gershoni}

\affiliation{The Physics Department and the Solid State Institute, Technion-Israel
Institute of Technology, Haifa 32000, Israel}

\begin{abstract}
Semiconductor quantum dots are considered to be the leading venue for fabricating on-demand sources of single photons. However, the generation of long-lived dark excitons imposes significant limits on the efficiency of these sources. We demonstrate a technique that optically pumps the dark exciton population and converts it to a bright exciton population, using intermediate excited biexciton states. We show experimentally that our method considerably reduces the DE population while doubling the triggered bright exciton emission, approaching thereby near-unit fidelity of quantum dot depletion. 
\end{abstract}
\maketitle

On-demand sources of single photons are an essential component of
emerging quantum information technologies \cite{knill2001,monroe2002,ladd2010,imamoglu1999, gisin2002, kimble2008}.  Semiconductor quantum dots (QDs) are a leading technology for achieving these single photon
emitters \cite{dekel2000, michler2000,santori2001,regelman2001,yuan2002,lounis2005, buckley2012}. In QDs, the emission of single photons results 
from recombinations of excited QD confined electron-hole pairs
(excitons) \cite{dekel2000, michler2000,santori2001,regelman2001}. The QDs are excited non-resonantly either optically \cite{santori2001,regelman2001} or electrically \cite{yuan2002, heindel2010} by a short pulse which leaves,  after relaxation, a confined exciton in the QD. 
In general, during the relaxation, the spins of the exciton pair are randomized such that they are either anti-parallel or parallel. An anti-parallel spin pair forms a bright exciton (BE), which can efficiently recombine optically and emit a single photon. In contrast, a parallel spin pair forms a long lived ($\sim$ 1 $\mu$s) dark exciton (DE) \cite{poem2010,mcfarlane2009, schwartz2015}, which cannot efficiently decay radiatively, preventing the device from being an on-demand emitter. 

In the case of resonant  excitation of the QD, resonantly or even quasi-resonantly tuned optical pulses can deterministically generate a bright exciton in the QD, providing an on-demand triggered source of single photons~\cite{muller2007,he2012} and even entangled photon pairs~\cite{akopian2006,muller2014}. Moreover, resonant optical excitations can be used to deterministically convert the light polarization into
exciton spin polarization~\cite{benny2011prl}. However, in practice, these resonant excitations require that the QD be totally depleted from charges and excitons prior to the exciting pulse.  
On-demand single photon QD-based emitters require a method to deterministically deplete the QD of dark excitons.    

Here, we present a method to deplete a QD of long-lived dark excitons. We optically pump the dark exciton population to the bright exciton population using intermediate excited biexciton states and we demonstrate experimentally both a substantial reduction in the DE population and a doubling of the triggered BE emission, thus proving depletion with close to unit fidelity. 

Excited biexciton levels consist of two electron-heavy hole
pairs where at least one charge carrier is in an excited energy level.
These levels have been the subject of several previous studies which provide a comprehensive understanding of their spectrum, spin wavefunctions of carriers in these levels, and their 
dynamics~\cite{poem2007, poemprb2010, benny2011prb,kodriano2010,schmidgall2014}.
Particularly relevant to this study are the nine states in which both the electrons and the heavy holes form 
spin-triplet configurations~\cite{kodriano2010,benny2011prb,schmidgall2014}.
Such states will be indicated in this paper by the notation $T^{e}_{m}$ ($T^{h}_{m}$) representing an electron spin triplet (hole spin triplet) of projection $m$ on the QD growth direction. We previously used two of these levels for demonstrating full coherent control of the 
bright exciton spin state ~\cite{poem2011,kodriano2012}. 

\begin{figure}[tb]
\includegraphics[width=8.5cm]{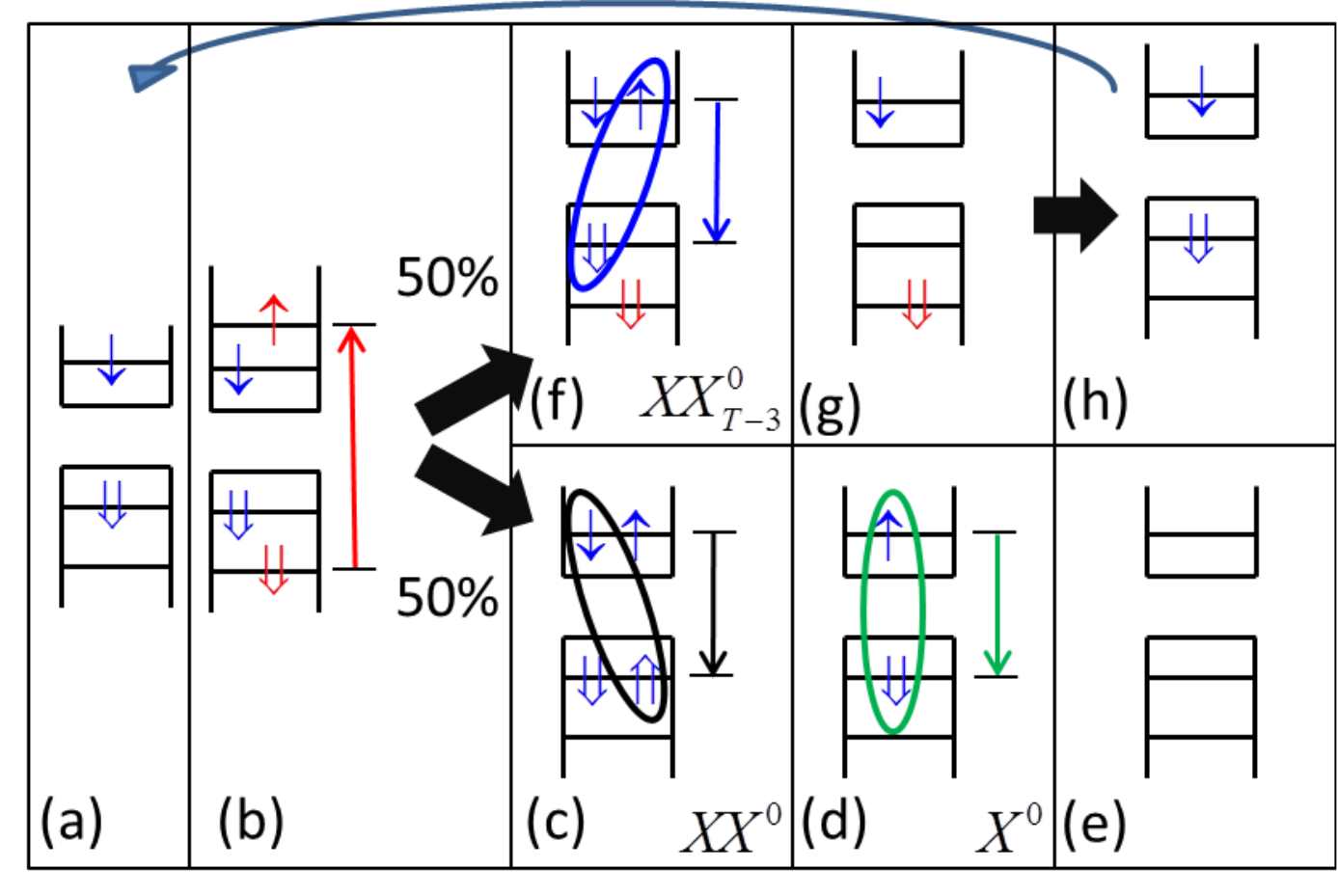}

\protect\caption{Schematic description of the optical depletion process. $\uparrow$($\downarrow$) [$\Uparrow$ ($\Downarrow$)] represents a spin-up (spin-down) electron [hole]. Blue (red) color is used for describing a ground (excited) state carrier. Upward (downward) arrows describe optical excitation (emission).
 (a) Following a non-resonant excitation pulse, the QD is populated with a long lived DE (only one spin projection is shown for clarity). (b) Optical excitation generates an excited
biexciton state. The biexciton then relaxes nonradiatively with almost equal probabilities either to its ground state $XX^0$ (c) or to a spin blockaded metastable state $XX^{0}_{T_{\pm 3}}$ (f). In the first case, the biexciton gives rise to a radiative cascade (c-d) which leaves the QD empty (e). In the second case, one photon is emitted (f) and after hole relaxation (g) the QD remains with a dark exciton (h). The process then repeats from (a), as long as the optical excitation lasts. Black arrows represent nonradiative processes. Radiative recombination is indicated by oval-matching
the recombining electron-hole pair, and oval colors are matched to the arrows indicating corresponding emission lines in Figure \ref{fig:PLEcw}(a).  \label{fig:schematic}}

\end{figure}

\begin{figure*}[htbp]
\includegraphics[width=12.8cm]{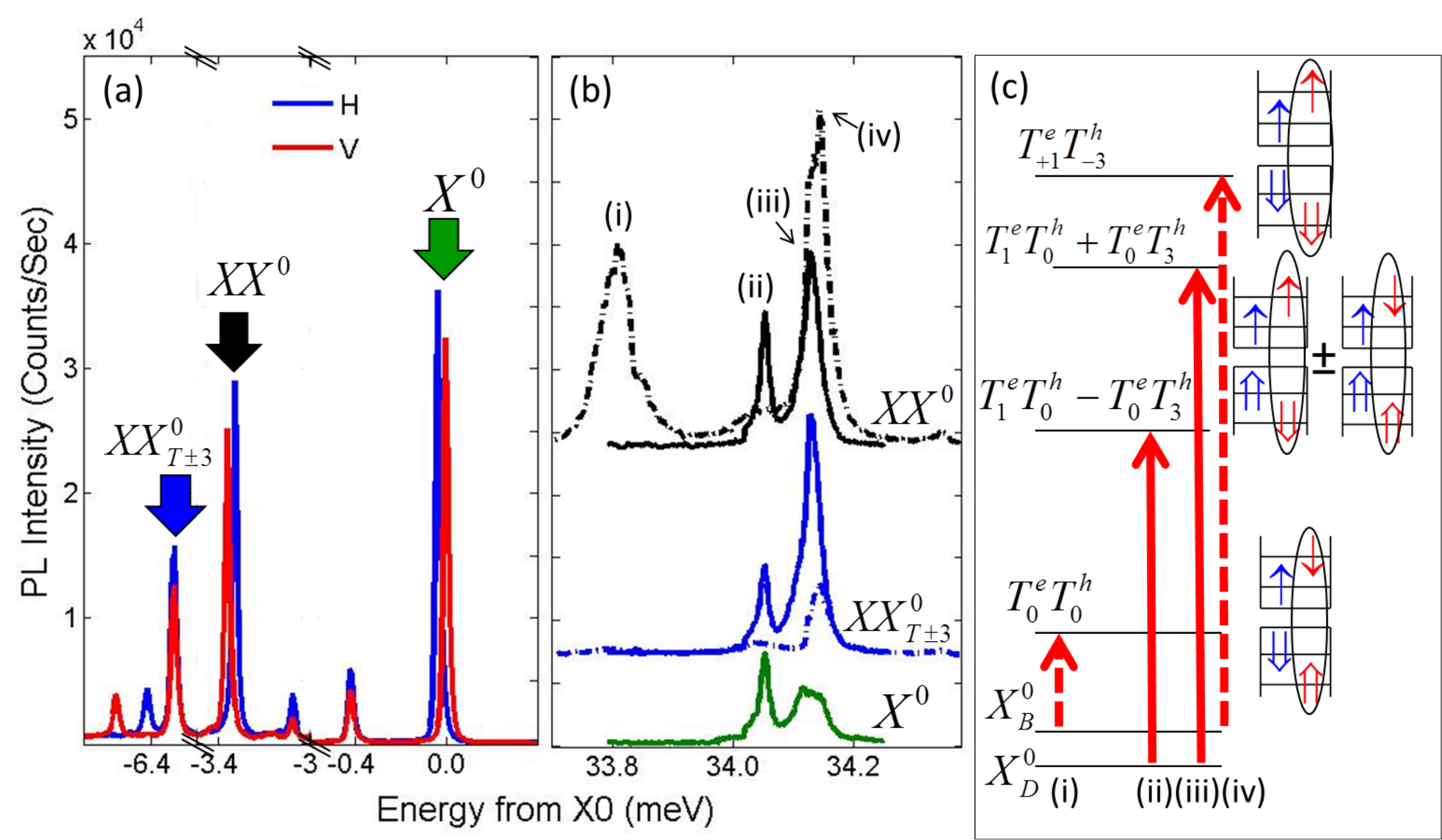}

\protect\caption{(a) PL spectra showing horizontally (H, blue) and vertically
(V, red) linearly-polarized emission from the QD. Relevant emission
lines are marked above the spectral line by the initial state of the optical transition.
Arrow colors match the PLE spectra in (b), which were obtained while monitoring the emission from the indicated spectral lines.
(b) PLE spectra of the indicated lines in a) showing the e-triplet h-triplet biexciton resonances relevant for the optical depletion. 
The PLE spectra were obtained using two pulsed lasers. Spectra indicated by solid (dashed) lines were obtained with one laser tuned to a dark (bright) exciton resonance while the energy of the second laser was scanned and the emission from the color matched PL line was monitored. 
(c) Schematic description of the optical transitions observed in (b). Optical transitions from the DE $X^{0}_{D}$ (BE, $X^{0}_{B}$) are indicated by solid (dashed) arrows and the added carriers are indicated in the spin state diagram by an oval-matched pair. The spin configuration of the state is provided on the left, where $T^{e}_{m}$ ($T^{h}_{m}$) represents an electron spin triplet (hole spin triplet) with projection $m$ on the QD $\hat{z}$ axis.  Roman numerals match the observed resonances in (b). Resonances labeled (ii) and (iii), which initiate from the DE and equally contribute to the DE and BE populations, can be used for the optical depletion process outlined in Figure \ref{fig:schematic}.
  \label{fig:PLEcw}}
\end{figure*}

In general, the nine $T^{e}_{m}T^{h}_{n}$ $m=-1,0,1, n=-3,0,3$ excited biexciton
states are spin blockaded from relaxation to the ground
e-singlet/h-singlet ($S^{e}_0S^{h}_0$) biexciton state~\cite{benny2011prb}.  There are, however, efficient spin flip and spin flip-flop 
mechanisms that permit this relaxation \cite{poemprb2010,benny2014, schmidgall2014}. In these processes, an electron or an electron and a hole flip their
spins due to the enhanced effect of the electron-hole exchange
interaction in the presence of a near resonant electron-longitudinal
optic (LO) phonon Fr{\"o}hlich interaction~\cite{benny2014}. 

The process of optical depletion is schematically described in Figure \ref{fig:schematic}, where Figure \ref{fig:schematic}(a) 
describes a single DE in its ground state in the QD.  The DEs are generated by non-resonant optical or electrical excitation, where electron-hole pairs with high excess energy are generated in the vicinity of the QD.  During their relaxation the carrier spins are randomized, resulting in a stochastic QD excitonic population that is composed on average of approximately 50\% BEs and 50\% DEs. Here, we used a weak continuous wave (cw) 445 nm pulse of a few nanoseconds duration to excite the QD and generate a mixed excitonic population in the QD. Within 2-3 ns of the end of the non-resonant pulse, the BE population has recombined radiatively, resulting in an empty QD. On average, this occurs in about 50\% of the cases, corresponding to the cases in which the QD was populated with a BE. In the remaining cases, a long-lived DE remains in the QD for long after the end of the non-resonant excitation. 
In order to deplete the DE, an optical resonant excitation~\cite{poem2010, schwartz2015} into an excited biexcitonic level is used as described in Figure \ref{fig:schematic}(b).
These excited biexciton states then quickly relax non-radiatively to one of the lower energy levels of the biexciton as described in Figure \ref{fig:schematic}(c).

In roughly half of the cases (as shown in Figure \ref{fig:schematic} and experimentaly demonstrated in Figure \ref{fig:PLEcw}(b), below) , the relaxation is into the spin blockaded biexciton
$XX^0_{T_{\pm3}}$ states, where further relaxation is prohibited by the spin-parallel
configuration of the heavy holes ($T^{h}_{\pm3}$ - Figure \ref{fig:schematic}(f)). This relaxation proceeds
by emission of another photon and a return to a QD containing a DE~\cite{poem2010, schwartz2015}. 
In the other cases, the relaxation is to the ground biexciton level from which the 
biexciton decays by the well studied two photon radiative cascade~\cite{gammon1996, akopian2006, muller2014}, leaving the QD eventually empty of charges. 

The relatively large branching ratio (approximately 0.5) between the two processes provides an extremely efficient way of depleting DE populations from the QD. It can be easily estimated that, for a measured exciton (biexciton) radiative lifetime of 470 (270) ps \cite{schwartz2015}, the QD can be fully emptied using an excitation pulse of several nanoseconds. 

The sample that we study was grown by molecular-beam epitaxy on a
(001)-oriented GaAs substrate. One layer of strain-induced InGaAs
QDs \cite{garcia1997} was grown in a planar microcavity \cite{ramon2006}.
The measurements were carried out in a $\mu$-PL setup at 4.2 K. More details on the sample and the experimental setup can be found in previous publications \cite{poem2011, kodriano2014}.
Few-picosecond pulsed excitation, such as the pulses used to probe QD population, was performed
using a frequency-tunable, cavity-dumped dye laser pumped by a frequency-doubled
Nd:YVO$_{4}$ (Spectra Physics Vanguard$^{TM}$) laser. For these experiments, the cavity-dumped dye laser pulse rate was 9.5 MHz. The longer depletion pulse used grating-stabilized tunable diode laser emission modulated by an electro-optic
modulator (EOM). This EOM was synchronized to the pulsed dye lasers, and it permitted variable pulse 
duration with rise and fall times of less than half a nanosecond.  For the non-resonant excitation pulse, a 445nm diode laser modulated by a fast accousto-optic modulator was used to provide a weak excitation pulse with a stochastic population of BEs and DEs in the QD.

Figure \ref{fig:PLEcw} presents the (a) PL and (b) two laser  PL excitation
(PLE) spectra of the QD ~\cite{benny2011prb}. In Figure \ref{fig:PLEcw}(a), relevant
emission lines corresponding to optical recombination from the ground
state exciton ($X^{0}$), ground state biexciton ($XX^{0})$ and the  
relevant spin-blockaded biexciton ($XX^0_{T_{\pm3}}$) are indicated by colored arrows. The energy is measured from the BE ($X^0$) spectral line at 1.283~eV.
The spin wavefunctions of the initial biexciton states, and the final exciton states that give rise to these transitions, were described 
in previous publications~\cite{kodriano2010,poem2010,schwartz2015}. 
We note here that optical transitions from the ground state biexciton $XX^0$ and the $m=0$ spin-blockaded biexciton
$XX^0_{T_{0}}$ result predominantly in BEs and lead to sequential emission of a photon due to BE recombination, while 
the $XX^0_{T_{\pm3}}$ biexciton line heralds the presence of a DE in the QD.  
The arrow color corresponds to the monitored PL spectral line used to obtain the PLE spectra in Figure
\ref{fig:PLEcw}(b). 

\begin{figure}[tb]
\includegraphics[width=8.5cm]{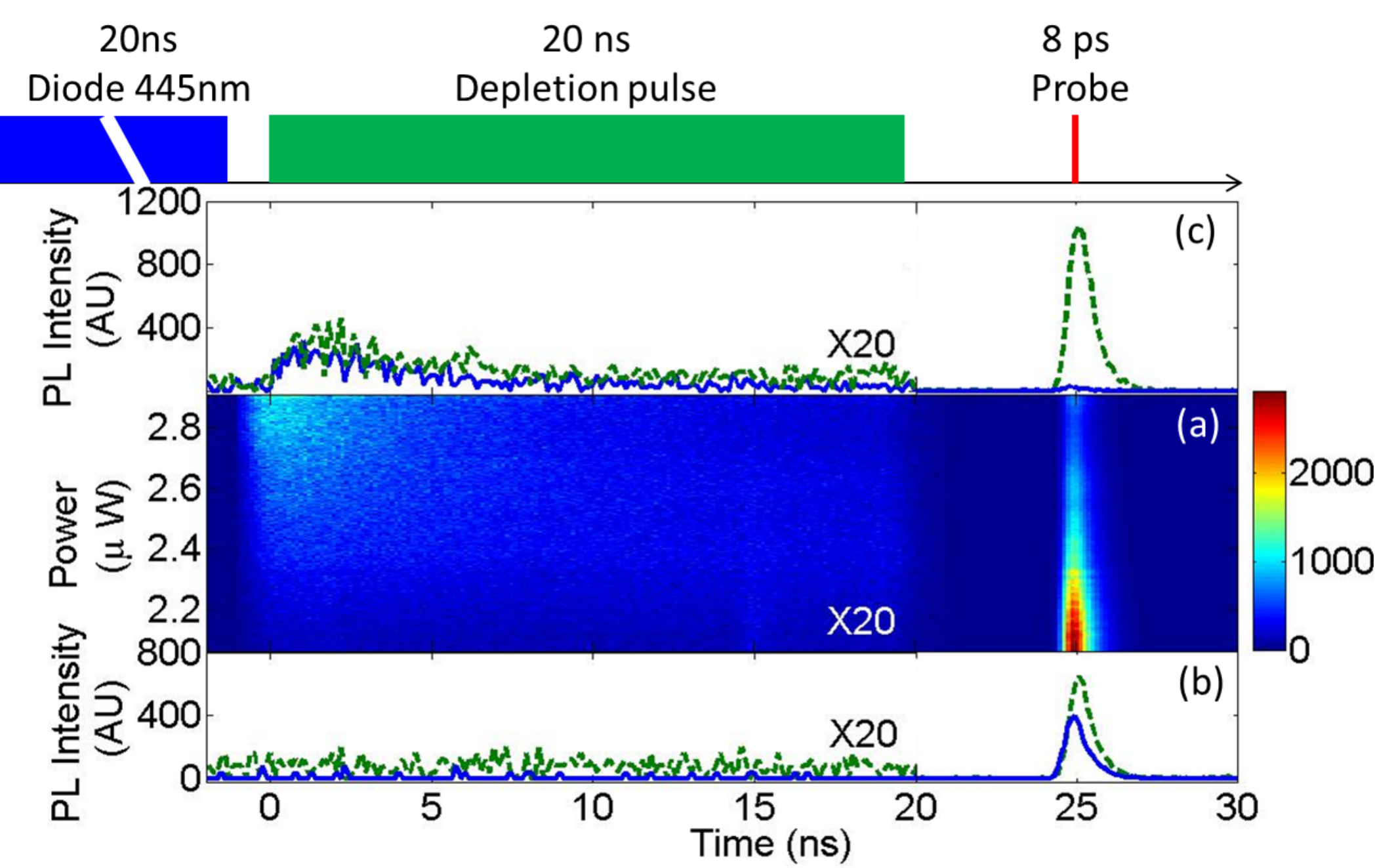}

\protect\caption{ (a) PL emission intensity from the $XX^0_{T_{\pm3}}$ spectral line (as given by the color bar) as a function of the power of the depleting pulse to the (iii) resonance  (Figure  \ref{fig:PLEcw}) and time. The temporal sequence of the non-resonant excitation pulse, depleting pulse, and the probe pulse tuned to the $XX^0_{T_{\pm3}}(X^{0})$ absorption line are shown above the Figure.
(b) [(c)] The solid blue line describes the $XX^0_{T_{\pm3}}$ PL emission intensity as a function of time at very low [maximal] depleting pulse power. More than 95\% depletion is clearly observed at high power. The overlaid dashed green line presents the intensity of the PL from the $X^{0}$ spectral line while the ps probe pulse is tuned to the BE resonance. The BE PL increases between (b) and (c) by nearly a factor of 2, yet another indication for the efficient depletion of the QD from DEs. \label{fig:signal}}

\end{figure}

Figure \ref{fig:PLEcw}(b) presents cw excited PLE measurements. One cw laser
is used to excite the BE (dashed lines) or DE (solid lines) \cite{schwartz2015}.
The energy of the second laser is varied while the PL emission
from the indicated emission line is monitored. Four exciton-biexciton resonances marked by Roman numerals are observed. 
Resonances indicated (i) and (iv) initiate from the BE, while resonances indicated (ii) and (iii) initiate from the DE. These optical transitions, which were previously
discussed~\cite{benny2011prb, schmidgall2014}, are schematically presented in Figure \ref{fig:PLEcw}(c).

Figure  \ref{fig:PLEcw} clearly demonstrates that while the resonances (i) and (iv), which initiate from the BE, contribute mainly to BE emission, resonances (ii) and (iii), which initiate from DE (solid lines),  contribute almost equally to the $XX^{0}_{T\pm3}$ emission and to the $XX^{0}$ emission. This means that in about 50\% of these excitations the DE population is transformed into a BE population and depleted, by subsequent optical recombination, from the QD. We chose resonance (iii) for the optical depletion outlined in Figure \ref{fig:schematic} above, though resonance (ii) performs similarly. 

For demonstrating optical depletion, a sequence of three laser pulses was used, schematically illustrated above Figure \ref{fig:signal}.
The first pulse was a weak 20 ns pulse duration of 445nm laser light.
A few ns after the pulse ended, a second pulse of 20 ns duration tuned to resonance (iii) of Figure  \ref{fig:PLEcw} was launched. The third pulse was an 8 ps long pulse, and it was tuned to the $XX^0_{T_{\pm3}}$  absorption line. It was turned on 5 ns after the depletion pulse ended, its intensity corresponded to a $\pi$-pulse, and the PL emission that resulted from this pulse monitored the population of the DE in the QD.

Figure \ref{fig:signal}(a) presents the PL intensity from the $XX^0_{T_{\pm3}}$ line as given by the color bar, as a function of time (horizontal axis) and the power of the depleting pulse tuned to the (iii) resonance of Figure  \ref{fig:PLEcw}(c) (vertical axis).  As can be readily seen in Figure \ref{fig:signal}(a), as the power of the depleting pulse increases, the population of the DE in the QD decreases.
A quantitative measure of the efficiency of the depleting pulse is provided by comparing the PL intensity during the probe pulse (solid blue line) in Figure \ref{fig:signal}(b), where the depletion pulse power was very weak (lower part of Figure \ref{fig:signal}(a)) to that in (c) where the power of the depletion pulse is maximal (upper part of Figure \ref{fig:signal}(a)). The integrated PL intensity during the probe pulse is reduced under these conditions to less than 5\%.

A complementary verification of the efficiency of the depletion is provided by comparing the PL emission from the BE (dashed green line in Figure \ref{fig:signal}(b) and (c)), at low (b) and high (c) depletion pulse power. For these measurements the probe pulse was tuned to one of the BE resonances \cite{benny2011prb}. The integrated BE emission during the probe pulse increases by almost a factor of 2, indicating that the QD is approximately 50\% occupied by DEs in the absence of the depletion pulse and predominantly empty when the depletion pulse is present. Under these conditions, we were able to get more than 8500 counts/sec on the BE detector during the probe pulse. Taking into account the repetition rate of our experiment (9.5MHz) and the measured overall light harvesting efficiency of the sample and the setup (approximately $1/1000$ \cite{schmidgall2014}) one gets depletion fidelity of close to 90\%.
We note here that, by increasing the depletion pulse power, we were able to achieve similar depletion fidelities with pulses as short as 3 ns (not shown).       

In conclusion, we have demonstrated an all-optical depletion method of semiconductor QDs which deterministically converts dark excitons to bright excitons, thereby increasing the fidelity of these on-demand triggered light sources. We do this using optical pumping via intermediate excited 
biexciton states which efficiently transfer dark exciton populations to bright excitons and subsequently to an empty QD, on a nanosecond timescale.
This fast depletion mechanism is essential for on-demand sources of single photons as well as for deterministic writing of the bright~\cite{benny2011prl} and dark exciton~\cite{schwartz2015} spin states using picosecond optical pulses. 
These abilities may be important for further development of QD-based devices for quantum information and communication applications. 

\begin{acknowledgments}
The support of the Israeli Science Foundation (ISF), 
the Israeli Nanotechnology Focal Technology Area on "Nanophotonics for Detection" 
and the German-Israeli Foundation (GIF) are gratefully acknowledged. 
\end{acknowledgments}

\bibliographystyle{unsrtnat}
\bibliography{depletion_apl_bib}

\end{document}